\documentclass[submission]{eptcs}
\usepackage{graphicx} 
\usepackage{amsmath,amsfonts,amssymb}
\usepackage{amsthm}
\usepackage{xcolor}
\usepackage{graphicx}
\usepackage{tikz,pgf}
\usetikzlibrary{automata,arrows,shapes,decorations,topaths,trees,backgrounds,shadows}
\usepgfmodule{matrix}
\usepackage[all]{xy}
\SelectTips{cm}{}

\renewcommand{\phi}{\varphi}
\newcommand\DUnion{
  \mathop{\mathchoice
    {\ooalign{$\displaystyle\bigcup$\cr\hss\scalebox{.65}{\raisebox{0.45ex}{$\bullet$}}\hss}}
    {\ooalign{$\textstyle\bigcup$\cr\hss\scalebox{.8}{\raisebox{0.5ex}{\tiny$\bullet$}}\hss}}
    {\ooalign{$\scriptstyle\bigcup$\cr\hss\scalebox{.45}{\raisebox{0.3ex}{$\bullet$}}\hss}}
    {\ooalign{$\scriptscriptstyle\bigcup$\cr\hss\scalebox{.38}{\raisebox{0.4ex}{$\bullet$}}\hss}}
    }
}

\title{Common Knowledge, Sailboats, and Publicity }
\def\titlerunning{CK, Sailboats, and Publicity}
\def\authorrunning{Bozdag \& Roy}
\author{Sena Bozdag\footnote{Universit\"{a}t Bayreuth} \and Olivier Roy\footnote{Universit\"{a}t Bayreuth}}

\begin{document}

\maketitle

\begin{abstract}
We revisit a recent puzzle about common knowledge, the ``sailboat" case \cite{Led2018uncommon}, and argue that Lewisian common knowledge allows us to reconcile the pre-theoretical intuition that certain facts are ``public" in such situations, while these facts cannot be common knowledge in the classical, iterative sense. The crux of the argument is to understand Lewisian common knowledge as an account of what it means for an event to be public. We first formulate this argument informally to clarify its philosophical commitment and then propose one way to capture it formally in epistemic-plausibility models. Taken together, we take the philosophical and the formal arguments as providing evidence that Lewisian common knowledge is a plausible account of what it means for an event to be public.
\end{abstract}
\noindent 

\noindent The various mathematical definitions of common knowledge, most prominently the iterative and fixed point definitions~\cite{barwise1988three,Bar89}, have often been associated, explicitly or implicitly, with the pre-theoretical notions of something being ``public information", ``out in the open", or what ``any fool"~\cite{mccarthy1978model} would know. Barwise's~\cite{barwise1988three} argument that ``the fixed point is the best conceptual analysis of the pre-theoretic notion" implicitly takes the latter to be some form of publicity.  Geanakoplos' early survey \cite{geanakoplos1994common} explicitly states that ``public events are the most obvious candidates for common knowledge." Similar remarks can be found in~\cite{milgrom1981axiomatic,clark1981definite} and in informal presentations of the characterizations of iterative common knowledge in terms of ``self-evident events"~\cite{geanakoplos1994common,osborne1994course}. This association between the pre-thereotical notion of publicity and the mathematical definitions of common knowledge is also salient outside epistemic logic and game theory, for instance, in philosophy of collective action~\cite{gilbert1992social,bratman2013shared,blomberg2016common}, or in Chwe's work on social rituals~\cite{chwe2013rational}.  

Many of the classical and more recent criticisms of the iterative or the fixed-point definitions of common knowledge can be diagnosed as questioning this association, c.f.~\cite{paternotte2011being} for an overview. A common criticism is, for example, that iterative common knowledge is unattainable for resource-bounded agents due to its infinitary character~\cite{parikh2005logical}. This fact is most worrying when one assumes that information can be public among such agents, and that iterative common knowledge should somehow capture this notion of publicity. One can similarly diagnose the ``Halpern-Moses problem"~\cite{Hal,morris2014coordination,gonczarowski2023common}. The paradoxical nature of the problem is most salient when one grants, on the one hand, that successful asynchronous coordination often rests on the publicity of some relevant facts and, on the other hand, that the iterative or the fixed-point definition of common knowledge should capture the latter.

A natural response to these worries, which we follow in this paper, is to propose an alternative account of publicity and argue that it meets some of the challenges facing the iterative or fixed-point definitions of common knowledge~\cite{gilbert1992social,gilbert2007mutual,williams2023publicity}. Here we follow a suggestion made by Paternotte~\cite{paternotte2011being} to use \emph{Lewisian} common knowledge~\cite{lewis1969convention,Cub,sillari2008common,vromen2023reasoning} to flesh out the pre-theoretical understanding of publicity. While Paternotte provides general, philosophical arguments for understanding publicity in terms of Lewisian common knowledge, in this paper we pitch this suggestion against a structurally related ``paradox" of iterated common knowledge: Lederman's ``Sailboat" example~\cite{Led2018uncommon}. In doing so, we argue that Lewisian common knowledge allows us to reconcile the pre-theoretical intuition that some obvious facts are public with the observation that iterative common knowledge of these facts do not hold, even for ideal reasoners.

The rest of this extended abstract is organized as follows. After briefly introducing the iterated and the Lewisian definitions of common knowledge, we sketch Lederman's Sailboat example and our philosophical argument to the effect that Lewisian common knowledge addresses it satisfactorily. We then provide a formalization of our argument in a variant of epistemic-plausibility models~\cite{baltag2008qualitative,van2011logical}. Section~\ref{sec:conclusion} concludes that Lewisian common knowledge constitutes a plausible account of publicity.

\section{Iterated and Lewisian Common Knowledge}
For the present paper, only the iterative definition of common knowledge is relevant. We use its standard version: A proposition $A$ is shared knowledge in a group $G$ whenever every member of $G$ knows $A$. $A$ is level-$n+1$ shared knowledge in group $G$ whenever it is shared knowledge in $G$ that $A$ is level-$n$ shared knowledge. Taking plain shared knowledge as level $0$, iterative common knowledge of $A$ is defined as level-$n$ shared knowledge of $A$, for all $n \in \mathbb{N}$. 

Lewisian common knowledge differs from the iterative definition in two ways. First and foremost, despite its name, it does \emph{not} feature higher-order knowledge, but rather generates higher-order \emph{reasons to believe}.\footnote{Vromen~\cite{vromen2023reasoning} provides textual evidence that David Lewis himself thought \emph{ex post} that ``common knowledge" was a misnomer for the notion he defined.} Second, Lewisian common knowledge requires the existence of a specific \emph{state of affairs}, the ``basis" for common knowledge, which grounds the group members' higher-order reasons to believe. In that sense, Lewisian common knowledge has a much more ``externalist" flavor~\cite{paternotte2011being}, since it refers to shared aspects of the informational environment~\cite{Bar89}, instead of referring to the members' mental states. 

We use Cubitt and Sugden's~\cite{Cub} rendering of the Lewisian definition. Let $R_i(...)$ stand for ``$i$ has a reason to believe...". A proposition $B$ is then said to be \textit{Lewisian common knowledge} in group $G$ whenever there exists a state of affairs $A$ such that:
\begin{description}
    \item[C1] If $A$ then $R_i(A)$ for any agent $i$
    \item[C2] For all agents $i,j$ if $R_i(A)$ then $R_iR_j(A)$
    \item[C3] For any agent $i$,  if $R_i(A)$ then $R_i(B)$
    \item[C4] For all agents $i,j$ and proposition $C$,  from the fact that $R_i(A)$ implies $R_i(C)$ it follows that $R_i(R_j(A)$ implies $R_j(C))$ 
\end{description}
Conditions C1 to C3 encapsulate the idea that $A$ is the basis for common knowledge, or what Cubitt and Sugden call a reflexive common indicator of $B$. These conditions do most of the work in generating the infinite hierarchy of 
higher-order reasons to believe $B$. C4 states that the agents share reasoning standards.\footnote{Observe that $A$ is a state of affairs and not a proposition. $R_i(A)$ is thus somewhat an abuse of notation and should be read as ``$i$ has a reason to believe that $A$ holds." Note also that our formulation of C1-C4 slightly differs from Cubit and Sugden's original formalization. Theirs also uses the Lewisian notion of ``indication" in C2-C4. That notion is, however, defined in terms of conditional reasons to believe, and once unpacked it yields the formulation above.}

Lewis argued informally that these conditions together entail that every member of $G$ has iterated higher-order reasons to believe $B$, which we write $r^n(B)$ following Cubitt and Sugden's~\cite{Cub} notation. In other words, the existence of the relevant state of affairs $A$ satisfying C1 to C4 entails that both agents have a reason to believe $B$, they have a reason to believe that they have a reason to believe $B$, and so on. Lewis' argument, sometimes also called ``Lewis' Theorem," has been formalized using a variety of tools~\cite{Cub,sillari2008common,vromen2023reasoning}.

\section{The Sailboat Case}
Our goal in this paper is to provide support to the view that Lewisian common knowledge constitutes a plausible account of what it means for a proposition or a state of affairs to be public. We do so by analyzing the ``Sailboat case" devised by Lederman~\cite{Led2018uncommon}. Before explaining in more detail the structure of our argument, let us introduce the case.

\begin{quote}
    SAILBOAT: Roman and Columba are ideal reasoners playing in a game show.\footnote{For our argument the fact that Roman and Columba are competing in a game show is not important.}
Each contestant has a single button on a console in front of him or her.
They have an unobstructed view of each other’s faces, and of an area in the
middle of the stage, where the hosts will place a sailboat. First, the hosts
will bring out a toy sailboat (the `test’) with a 100 cm mast. They will then
replace it with a sailboat chosen randomly from an array of sailboats of
various sizes. If the mast of the new sailboat is taller than the test and both
players press their respective buttons, they receive \$ 1,000 each. If the mast
is not taller than the test and both press, or if only one person presses their
button, the person or people who pressed must pay the show \$ 100. Today,
the mast of the chosen boat is 300 cm tall. \cite[p.1075]{Led2018uncommon}
\end{quote}
\noindent Lederman assumes that the reader will judge the following to be intuitively true in this example, which we also accept for the sake of argument in this paper. 
\begin{enumerate}
    \item[]\label{pubthresh} It is public among Roman and Columba that the mast is larger than 100 cm. 
\end{enumerate}
We write $[>100]$ for ``the mast is larger than 100 cm." Here Lederman uses a pre-theoretical notion of publicity, which stays unanalyzed in the paper. Our point here is precisely that this notion can be fleshed out in terms of Lewisian common knowledge.

Lederman argues that, even though this proposition is public,  $[>100]$ is not common knowledge in the iterative sense. The argument rests on the assumption that Roman and Columbia's knowledge is inexact, that is, it has a margin of error $m$, and that this is public between them. In his words: \begin{quote}
[Interpresonal Ignorance] For all r, if it looks to be r cm tall to one of
the agents, then for all that agent knows, it looks to be $[r\pm m]$ cm tall
to the other. \cite[p.1081]{Led2018uncommon}
\end{quote}

In this quote, ``for all the agent knows" is capturing the notion of ``considering it possible", or ``not being ruled out by the agent's information" frequently represented by the ``diamond" operator in epistemic logic.  Interpersonal ignorance thus says that for all $n$, if it looks to an agent that the mast is $n$ cm tall, then that agent considers it possible that the mast looks $n-m$ tall to the other. 
That fact is also considered to be public. The argument that they do not have common knowledge then runs like many of the classical examples or ``paradoxes" of common knowledge, most prominently the Byzantine Generals case. It builds a chain of ``Roman considers it possible that Columba considers it possible that..."\footnote{With this in hand Lederman goes on to argue for the admittedly more thought provoking claim that, for all practical purposes, iterative common knowledge does not exists, because cases of inexact knowledge are overwhelmingly common. This is much stronger than what is suggested by the Sailboat case alone, and our argument in this paper is independent of whether that stronger claim is correct.}

\section{Philosophical Argument}\label{sec:philarg}
We now present a first version of our argument. We claim that even granting $[>100]$ is not common knowledge in the iterative sense, Lewisian common knowledge can recapture the pre-theoretical intuition that $[> 100]$ is public. In other words, a failure of iterative common knowledge of $[>100]$ is consistent with Lewisian common knowledge of that proposition. 

Three remarks are in order before proceeding. First, we start with an informal version of the argument to clarify its philosophical commitments. Second, despite us talking in terms of consistency, we do not take issue with Lederman's claim that theories associating publicity and iterative common knowledge for ideal reasoners are inconsistent with the fact that $[>100]$ is public in the Sailboat case~\cite[p.1075]{Led2018uncommon}. Our argument can, in fact, be read as adding support to that claim. Relatedly, it is worth re-emphasizing that we follow Lederman and assume that Roman or Columba are ideal reasoners. So the argument is not based on them only being able to entertain finitely many levels of knowledge or on them not forming the beliefs that they have reasons to form. The argument rests rather on the simple observation that having a reason to believe a proposition $A$ is consistent with considering $\neg A$ possible, i.e. not completely ruling out that $A$ might be false.

\paragraph{Assumption: existence of a reflexive common indicator $A$:} To argue that $[>100]$ is Lewisian common knowledge in the Sailboat example we need to identify a ``basis", i.e. the reflexive common indicator $A$ that grounds the relevant higher-order reasons to believe. 
We simply take for granted that because Roman and Columba ``have an unobstructed view of each other's faces and an area in the middle of the stage, where the hosts will place a sailboat" they are in a ``shared environment" that constitutes the relevant basis $A$ for Lewisian common knowledge. We furthermore assume that $A$ can be approximately described by a proposition, so that statements of the form ``$R_i(A)$" are meaningful. This means, in other words, that we assume that conditions C1 and C2 hold for this appropriately described $A$. Finally, we assume that this $A$ entails certain other propositions, crucially with regard to the margin of errors and $[>100]$. 

\paragraph{Assumption: shared standards of reasoning:} We also assume that Roman and Columba have ``shared standards of reasoning." They are ideal reasoners, and this is public between them. Standards of reasoning go, however, beyond ``mere" logical omniscience, since they also capture evidential relations and common sense, defeasible inferences that either of them can make. This should be modeled more precisely using some defeasible or nonmonotonic theories for the ``implies" and ``if... then" statements in C4, but we leave that issue aside in this paper. See, however, ~\cite{paternotte2011being} for one way to model defeasibility in Lewisian common knowledge, using probabilistic credences. 

\paragraph{Publicity of the margins:} Let us write $[n-m,n+m]$ for the fact that the mast is at least $n-m$ cm, at most $n+m$ cm tall, with $n$ being any number and $m$ the given margin of error. We write $[n]$ for the fact that the mast is exactly $n$ cm tall. Publicity of the margins would then correspond to saying that $A$ is a reflexive common indicator for the following: 

\begin{equation}
  \forall n \forall i \text{ if } [n] \text{ then } R_i([n-m,n+m])\label{marg}\tag{Margins}
\end{equation}
For this we need to argue that C1-C4 hold for~\eqref{marg}, given the basis $A$ that we assume exists. We have already mentioned that C1, C2, and C4 hold by assumption about $A$. The fact that the margins are public then simply boils down to saying that C3 holds with $B$ being~\eqref{marg}. In other words, if Roman and Columba have a reason to believe that $A$ holds, then they have a reason to believe~\eqref{marg}. We find it a plausible assumption.

By ``Lewis' Theorem" we then obtain that there is iterated, higher-order reason to believe~\eqref{marg}, which we write $r^n(\ref{marg})$.
 
In particular, for every finite level $k$, we have that Roman and Columba have level-$k$ reason to believe that~\eqref{marg} holds. For instance we get that:

$$R_{R}(\forall n \forall i \in \{R, C\} \text{ if } [n] \text{ then } R_i([n-m,n+m]))$$
$$R_{C} R_{R} (\forall n \forall i \in \{R, C\} \text{ if } [n] \text{ then } R_i([n-m,n+m]))$$
$$R_{R}R_{C} R_{R} (\forall n \forall i \in \{R, C\} \text{ if } [n] \text{ then } R_i([n-m,n+m]))$$
... and so on, where $R_{R}(...)$ should be read as ``Roman has a reason to believe that...", and similarly for $R_C$ and Columba. We furthermore assume that reasons to believe are sufficiently closed under logical consequence\footnote{In the philosophical argument we require closure under logical consequence only to go from the general form of $r^n(\ref{marg})$ to its instantiations. In the formal argument, however, use the standard semantics for the modal operators, which means that both agents are modeled as logically omniscient.} for Roman and Columba to have the corresponding (higher-order) reasons to believe every instantiation of these conditionals with specific values of $n$ and agents $R$ or $C$. We assume, for instance, that the following also holds:
$$R_{R}(\text{ if } [300] \text{ then } R_C([300-m,300+m]))$$
$$R_{C} R_{R} (\text{ if } [300-m] \text{ then } R_C([300-2m,300]))$$
$$R_{R}R_{C} R_{R} (\text{ if } [300-2m] \text{ then } R_C([300-3m,300-m]))$$
... and so on.

\paragraph{Publicity of $[>100]$:} We also assume that the basis $A$ for Lewisian common knowledge is such that C3 holds for $B$ being $[>100]$. So we also have $r^n([>100])$. Depending on the size of the margin, there must thus be a number $k$ of iterations of $R_C R_R ...$, which we write $r_k...$, and a number $l$ such that $300-(l+m) < 100$ and for which the following two statements hold.
$$r_k (\text{ if } [300-l] \text{ then } R_C([300-(l+m),(300-l)+m])$$ 
$$r_k (R_C[>100])$$
We now informally claim that these two statements are consistent. The first statement in the scope of $r_k$ is a conditional, and only its consequent contradicts $R_C([>100])$. As long as this consequent does \emph{not} detach, there is no inconsistency. It would detach if we had $r_k([300-l])$, assuming that the $R_i$ operators satisfy a version of the K axiom, and taking into account the type of defeasible conditional in the scope of $r_k$. We content that Roman and Columba do not have this type of unconditional iterated reason to believe, except for 300cm, the actual size of the mast. The epistemic-plausibility model that we construct below can be seen as a way to flesh out this part of the informal argument.

\paragraph{Consistency of publicity with failure of iterative common knowledge:}
Up to now our argument has been aimed at showing that even though the margins are public, it can be Lewisian common knowledge that the mast is greater than 100 cm. We now want to argue that this Lewisian common knowledge is consistent with Roman and Columba \emph{not} having iterative common knowledge of that same fact.  

We grant for the sake of the argument that Roman and Columba consider it possible that the mast is $300-m$. They have, after all, a reason to believe that $[300-m,300+m]$. As Lederman argues, because the margins of errors are public, both consider it possible that the other has a reason to believe that $[300-2m,300]$, which in turn entails that they both consider it possible that the other considers it possible that the other has a reason to believe that $[300-3m,300-m]$, and so on. This iteration does not stop when the lower bound of the interval is below 100, so it is not common knowledge between Roman and Columba, in the iterative sense, that the mast is taller than 100 cm.

This failure of iterative common knowledge of $[>100]$ is consistent with Lewisian common knowledge of that proposition for the simple reason that not ruling out something, say $\neg C$, is consistent with having a reason to believe that $C$ is, in fact, the case. Recall that we follow Lederman and assume that Roman and Columba both consider it possible that the mast is $300-m$ cm tall. This does not entail that either of them has a reason to believe $[300-m]$, or has a reason \emph{not} to disbelieve $[300-m]$. Reasons to believe come in various strengths, and even a conclusive reason to believe might not ground absolute certainty. Roman and Columba might be quite confident in their assessment of the mast, but they are also aware of the margin of error. This means that even though the reason for $[300]$ might be the strongest piece of evidence that each has, that reason might not be strong enough to completely rule out $[300-m]$. In other words, having a reason to believe $[300]$ is perfectly consistent with not ruling out $[300-m]$. 

This point generalizes to all $[n]$, which allows the iteration of ``Roman considers it possible that Columba considers it possible that Roman..." to continue past the lower bound of 100 cm without threatening Lewisian common knowledge of $[>100]$. As we saw in the previous section, because the margins are public, Roman and Columba have a conditional reason to believe that \emph{if} the mast \emph{was} $300-m$ tall, then the other \emph{would} have a reason to believe $[300-2m, 300]$. And, of course, since they consider $[300-m]$ possible, they do consider it possible that the other has a reason to believe $[300-2m, 300]$ and, by the same argument as in the previous paragraph, this is consistent with the other considering it possible that the mast is $300-2m$ cm tall. But this, so to speak, ``possible reason to believe", does not translate into an unconditional, second-order reason to believe that the other has a reason to believe $[300-2m, 300]$, and \emph{a fortiori} that the other has a reason to believe that the mast is $300-2m$ cm tall. Each has, after all, a whole array of such possible reasons to believe, one for each number $n$ in the interval $[300-m,300+m]$. Consequently, while there is a number of iterations of ``Roman considers it possible that Columba considers it possible that..." for which the mast is less than 100 cm tall, because the consequents of the relevant conditional reasons to believe do not detach, neither at the first nor at any higher-order levels, this failure of iterated common knowledge does not clash with the Lewisian common knowledge of $[>100]$. Again, ``considering it possible" is different, and crucially much weaker than having a reason to believe.\footnote{This point can also be made using the distinction between pro tanto and conclusive reasons~\cite{broome2005does}. We omit it here for reasons of space, but we elaborate on it in the full version of the paper. }

\section{An epistemic-plausibility model to flesh out the philosophical argument}\label{sec:epm}
We now argue that the intuitions underlying our philosophical argument can be fleshed out using a variant of epistemic-plausibility models~\cite{baltag2008qualitative,van2011logical}. We should emphasize at the outset that we interpret these models in a slightly non-standard way, for two reasons. First, these models and the language describing them have been developed to study the relation between knowledge and conditional beliefs, and not necessarily the reasons supporting the latter. Second, the standard semantics of conditional beliefs in epistemic-plausibility models is not well suited to handle cases where the condition is known to be false. To address this, we augment epistemic plausibility models with selection functions, a standard tool for analyzing counterfactual beliefs in belief revision~\cite{alchourron1985logic} and game theory~\cite{stalnaker1996knowledge,halpern2001substantive}.

Before we proceed, a further disclaimer is in order: We do \emph{not} claim that the specific variant of epistemic-plausibility models we use provides, in general, a good account of reasons to believe or of Lewisian common knowledge, nor that it should be seen as an alternative to existing formalizations of the latter~\cite{Cub,sillari2008common,paternotte2011being,vromen2023reasoning}. In the full paper, we expand on the reasons why. Our claim is that, given the specific features and idealizations of the Sailboat case, the construction below allows us to flesh out the philosophical consistency argument in the previous section. 

\paragraph{Language:} We work with a propositional modal language whose propositional variables are of the form $[n]$, for all $n \in \mathbb{N}$, together with a designated proposition $\top$ that will denote propositional tautologies. We use $p$ to designate elements of $Prop=\{[n] : n \in  \mathbb{N} \} \cup \{\top\}$. Our set of agents $A$ is simply $\{\rho, \kappa\}$, for Roman and Columba. The propositional language is then extended with two modalities, $R_i(\phi|p)$ and $ K_i\varphi$, to be read ``agent $i$ has a reason to believe $\phi$, given $p$" and ``agent $i$ knows $\phi$." For technical reasons to be explained below, we restrict the syntax so that only propositional variables can appear as conditions in formulas of the form $R_i(\phi|p)$:

 $$p \ |\  \phi \wedge \psi \ |\ \neg \phi \ |\  R_{i\in A}(\phi | p)\ |\ K_{i\in A}\phi$$

Formulas of the form $R_i(\phi|p)$ express conditional reasons to believe. We have a subjective interpretation of these conditional reasons, such as ``$i$ has a reason to believe that, given $p$, $\phi$ is true", as opposed to ``given $p$, $i$ has a reason to believe $\phi$". Unconditional reasons to believe $R_i(\phi)$ are defined as $R_i(\phi|\top)$. These formulas have a dual, which we will write $\langle R_i \rangle (\phi | p)$ and define as $\neg R_i(\neg \phi|p)$, and $\langle K_i \rangle \varphi$, defined as usual. We do not use the dual of having reasons to believe in our argument below, whereas $\langle K_i \rangle \varphi$ is the object-language expression of ``considering it possible" modality we introduced early on. With this in hand, we are able to define the higher-order notions that we will need below: 
    \begin{itemize}
        \item $r^0(\phi | p) = \bigwedge_{i \in A} R_i(\phi|p)$. Given $r^k (\phi|p)$, we define $r^{k+1}(\phi | p)$ as $\bigwedge_{i \in A} R_i(r^{k}(\phi|p) | p)$, and $r^n (\phi | p)$ as $r^k(\phi |p)$, for all $k \in \mathbb{N}$.
        \item Similarly, $\langle K^0_i \rangle \varphi = \langle K_i \rangle \varphi$, and $\langle K^{k+1}_i \rangle \varphi$ as $\langle K_i \rangle \langle K^{k}_j \rangle \varphi$ for $i \neq j$, and $\langle K^{n}_i \rangle \varphi$ as $\langle K^{k}_i \rangle \varphi$ for all $n$.\footnote{For the construction below we need only this limited, diamond version of higher-order knowledge, but we take the iterative common knowledge to be defined in the usual way.}
    \end{itemize}

\paragraph{Epistemic-Plausibility Frames, Models, and Selection Functions:} An epistemic-plausibility frame $\mathcal{F}$, given our set of agents $A$  is a tuple $\langle W, \{\leq_i\}_{i \in A}\rangle$ where $W$ is a set of states and each $\leq_i$ is a preorder, i.e., reflexive and transitive relation on $W$. The strict relations $w <_i w'$ are defined as $w \leq_i w'$ but not $w' \leq_i w$. The \emph{epistemic accessibility} relation $\approx_i$ is defined as: $w \approx_i w'$ iff either $w \leq_i w'$ or $w' \leq_i w$. We write $[w]_i$ for $\{w' : w \approx_i w'\}$. Given our set of propositional letters $Prop = \{\top\} \cup \{[n]: n \in \mathbb{N}\}$, a \emph{epistemic-plausibility model} $\mathcal{M}$ is an epistemic-plausibility frame $\mathcal{F}$ together with a valuation $V: Prop \to \mathcal{P}(W)$ such that $V(\top) = W$. Given an epistemic-plausibility model $\mathcal{M}$, a \emph{selection function} for $\mathcal{M}$ is any $f: (Prop \times W) \to W$.\footnote{For the construction below we only need to define conditional reason to believe, and therefore selection functions, for the case where the condition is a propositional variable of our language. Hence the syntactic restriction on $R_i(\phi |p)$ and the definition of the selection function taking only arguments in $Prop$ instead of the full language $\mathcal{L}$.}

\paragraph{Truth conditions:} Let $\mathcal{M}$ be an epistemic-plausibility model and $f$ a selection function for it. Then we set:   
    \begin{itemize}
        \item  $\mathcal{M}, w \models p$ iff $w \in V(p)$.
        \item  $\mathcal{M}, w \models R_i(\phi|p)$ iff $max_{\leq_i}(||p|| \cap [f(p, w)]_i) \subseteq ||\phi||$.
        \item  $\mathcal{M}, w \models K_i \varphi$ iff $[w]_i \subseteq ||\phi|| $
    \end{itemize}
where $||\varphi|| = \{w : \mathcal{M}, w \models \varphi \}$ and for any $X \subseteq W$, $max_{\leq_i}(X)=\{w \in X : $ there is no $v \in X$ such that $w <_i v \}$.

\paragraph{Remarks:} The truth conditions for the two modal operators are spelled out using the underlying plausibility orderings $\leq_i$. This requires some explanation.

As their name suggests, the orderings $\leq_i$ are typically interpreted in terms of doxastic plausibility: $w \leq_i v$ is then meant to model the fact that $i$ considers $v$ at least as plausible as $w$. Here we slightly deviate from that standard interpretation. We take $w \leq_i v$ to model the fact that $i$ \emph{has a reason to believe} that $v$ is at least as plausible as $w$. We take these reasons to be subjective, conclusive and normative~\cite{mcnaughton2018motivating}. 

The conditional reason-to-believe operators are then interpreted using a combination of this plausibility ordering and the selection function. Leaving the latter aside for a moment, the semantics above yields that an agent has a reason to believe $\psi$ conditional on $p$ whenever $\psi$ is true in all $p$ states (compatible with $i$'s total evidence) for which $i$ has a reason to believe are maximally plausible.\footnote{This interpretation is not unlike standard semantics for conditional obligations in deontic logic that ``lift" reason-based orderings on states to orderings on propositions~\cite{jongh2009preference,dietrich2013reason}. Our treatment of conditional reasons to believe is similar also to treatment of counterfactual type of beliefs in the Lewis-Stalnaker tradition.}

The truth conditions for both the conditional reason to believe and the knowledge operators are relativized to the agent's informational environment $[w]_i$ at any given state. This environment is, in turn, defined using the plausibility orderings $\leq_i$: the equivalence relation $\approx_i$ generated by the connected components of $\leq_i$ \cite{baltag2008qualitative}. Philosophically, this gives the following interpretation. All states in $[w]_i$ are those that are not excluded for the reasons to believe that $i$ has at $w$. $[w]_i$ can thus be seen as capturing $i$'s total evidence at $w$ \cite{williamson1997knowledge}.

Returning to the selection functions, they are meant to capture $i$'s conditional reasons to believe $R_i(\phi |p)$ both in cases where $i$'s total evidence rules out the truth of $p$, and in cases where $i$'s total evidence is consistent with $p$. We define a concrete selection function below, but we could also define them using general constraints that characterize minimal revisions~\cite{stalnaker1996knowledge}. For example, one could require that if $f(\varphi, w) = w'$ then $\mathcal{M}, w' \models \varphi$, or that $f(\varphi, w) = w$ whenever $\mathcal{M}, w \models \varphi$. The selection function below satisfies the former, but not the latter.

\paragraph{Butterflies and the butterfly flutter:} We are now ready to describe the main construction that fleshes out the philosophical argument presented in the previous section. We construct a specific epistemic plausibility model that fulfills three functions. First, it describes the actual situation of Roman and Columba, where both the fact that the mast is 300 cm tall, and thus larger than 100 cm, and the margins of errors, are public. Here we capture publicity by the fact that Roman and Columba have iterated higher-order reasons to believe these propositions.\footnote{Although we argue indirectly for this, the formal argument thus leaves out an explicit representation of the fact that conditions C1-C4 hold for a given reflexive common indicator. The way publicity is captured in our formal model is therefore reminiscent of the understanding of publicity in terms of common commitment to believe~\cite{williams2023publicity}.} Second, the model captures the fact that they both entertain the relevant chains of ``Roman considers it possible that Columba considers it possible... that the mast is $k$ cm tall", even for $k$ smaller than 100. Finally, the model also captures the fact that Roman and Columba have the relevant reasons to believe that the first two conditions would hold, counterfactually, for any $k \neq 300$. The construction accordingly proceeds in three steps corresponding to these three functions.

Let $Prop$ and $A$ be as above. Now take $k,m \in \mathbb{N}$. Intuitively, $k$ will be the `center' of the `butterfly', and $m$ the margin, with $k - m > 0$. Our goal is to define inductively a \emph{butterfly centered on $k$} as an epistemic-plausibility model $\mathcal{M}_k$. We start with the base case. 

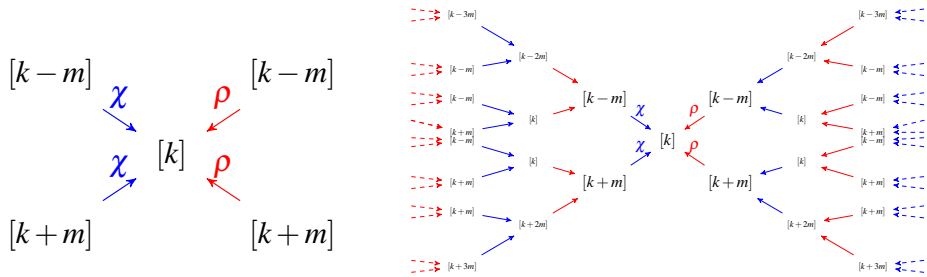
\begin{figure}[ht]
    \centering
      \resizebox{5cm}{!}{
    \begin{tikzpicture}[node/.style={circle, draw=black,minimum size=10mm, thick},->,>=stealth',shorten >=1pt,shorten <=1pt,
    auto, 
    node distance=2cm,
    ]
    \tikzstyle{every state}=[draw=none,text=black]
\node[state] at (0,0) (w0) {$[k]$};
\node[state] at (1.5,1) (w1) {$[k-m]$};
\node[state] at (1.5,-1) (w2) {$[k+m]$};
\node[state] at (-1.5,1) (w3) {$[k-m]$};
\node[state] at (-1.5,-1) (w4) {$[k+m]$};

 \draw[->, label,red] (w1) -- (w0) node [midway,above]{$\rho$};
  \draw[->, label,red] (w2) -- (w0) node [midway,above]{$\rho$};
  \draw[->, label,blue] (w3) -- (w0) node [midway,above]{$\chi$};
  \draw[->, label,blue] (w4) -- (w0) node [midway,above]{$\chi$};
    \end{tikzpicture}
    }
\resizebox{8cm}{!}{
    \begin{tikzpicture}[node/.style={circle, draw=black, thick},->,>=stealth',
    auto, 
    node distance=2cm,
    ]
    \tikzstyle{every state}=[draw=none,text=black]

\node[state] at (0,0) (w0) {$[k]$};
\node[state] at (1.5,1) (w1) {$[k-m]$};
\node[state] at (1.5,-1) (w2) {$[k+m]$};
\node[state] at (-1.5,1) (w3) {$[k-m]$};
\node[state] at (-1.5,-1) (w4) {$[k+m]$};

\draw[->, label,red] (w1) -- (w0) node [midway,above]{$\rho$};
  \draw[->, label,red] (w2) -- (w0) node [midway,above]{$\rho$};
  \draw[->, label,blue] (w3) -- (w0) node [midway,above]{$\chi$};
  \draw[->, label,blue] (w4) -- (w0) node [midway,above]{$\chi$};


\node[state] at (3.2,2) (w5) {\tiny{$[k-2m]$}};
\node[state] at (3.2,0.5) (w6) {\tiny{$[k]$}};

\node[state] at (3.2,-0.5) (w7) {\tiny{$[k]$}};
\node[state] at (3.2,-2) (w8) {\tiny{$[k+2m]$}};

\node[state] at (4.9,3) (w13) {\tiny{$[k-3m]$}};
\node[state] at (4.9,1.7) (w14) {\tiny{$[k-m]$}};

\node[state] at (4.9,1) (w15) {\tiny{$[k-m]$}};
\node[state] at (4.9,0.2) (w16) {\tiny{$[k+m]$}};

\node[state] at (4.9,0) (w17) {\tiny{$[k-m]$}};
\node[state] at (4.9,-1) (w18) {\tiny{$[k+m]$}};

\node[state] at (4.9,-1.7) (w19) {\tiny{$[k+m]$}};
\node[state] at (4.9,-3) (w20) {\tiny{$[k+3m]$}};

\node[state] at (6.6,3.2) (w21) {};
\node[state] at (6.6,2.8) (w22) {};

\node[state] at (6.6,1.9) (w23) {};
\node[state] at (6.6,1.5) (w24) {};

\node[state] at (6.6,1.2) (w25) {};
\node[state] at (6.6,0.8) (w26) {};

\node[state] at (6.6,0.6) (w27) {};
\node[state] at (6.6,0.2) (w28) {};

\node[state] at (6.6,0.1) (w29) {};
\node[state] at (6.6,-0.2) (w30) {};

\node[state] at (6.6,-0.8) (w31) {};
\node[state] at (6.6,-1.2) (w32) {};

\node[state] at (6.6,-1.5) (w33) {};
\node[state] at (6.6,-1.9) (w34) {};

\node[state] at (6.6,-2.8) (w35) {};
\node[state] at (6.6,-3.2) (w36) {};

   \draw[->, label,blue] (w5) -- (w1) ;
  \draw[->, label,blue] (w6) -- (w1);

   \draw[->, label,blue] (w7) -- (w2) ;
  \draw[->, label,blue] (w8) -- (w2);

  \draw[->, label,red] (w13) -- (w5) ;
  \draw[->, label,red] (w14) -- (w5);

   \draw[->, label,red] (w15) -- (w6);
  \draw[->, label,red] (w16) -- (w6);

  \draw[->, label,red] (w17) -- (w7);
  \draw[->, label,red] (w18) -- (w7);

  \draw[->, label,red] (w19) -- (w8);
  \draw[->, label,red] (w20) -- (w8);

   \draw[->, label,blue,dashed] (w21) -- (w13) ;
  \draw[->, label,blue,dashed] (w22) -- (w13);

  \draw[->, label,blue,dashed] (w23) -- (w14) ;
  \draw[->, label,blue,dashed] (w24) -- (w14);

  \draw[->, label,blue,dashed] (w25) -- (w15) ;
  \draw[->, label,blue,dashed] (w26) -- (w15);

\draw[->, label,blue,dashed] (w27) -- (w16) ;
  \draw[->, label,blue,dashed] (w28) -- (w16);

  \draw[->, label,blue,dashed] (w29) -- (w17) ;
  \draw[->, label,blue,dashed] (w30) -- (w17);

\draw[->, label,blue,dashed] (w31) -- (w18) ;
  \draw[->, label,blue,dashed] (w32) -- (w18);

  \draw[->, label,blue,dashed] (w33) -- (w19) ;
  \draw[->, label,blue,dashed] (w34) -- (w19);

   \draw[->, label,blue,dashed] (w35) -- (w20) ;
  \draw[->, label,blue,dashed] (w36) -- (w20);

\node[state] at (-3.2,2) (w9) {\tiny{$[k-2m]$}};
\node[state] at (-3.2,0.5) (w10) {\tiny{$[k]$}};

\node[state] at (-3.2,-0.5) (w11) {\tiny{$[k]$}};
\node[state] at (-3.2,-2) (w12) {\tiny{$[k+2m]$}};

\node[state] at (-4.9,3) (w37) {\tiny{$[k-3m]$}};
\node[state] at (-4.9,1.7) (w38) {\tiny{$[k-m]$}};

\node[state] at (-4.9,1) (w39) {\tiny{$[k-m]$}};
\node[state] at (-4.9,0.2) (w40) {\tiny{$[k+m]$}};

\node[state] at (-4.9,0) (w41) {\tiny{$[k-m]$}};
\node[state] at (-4.9,-1) (w42) {\tiny{$[k+m]$}};

\node[state] at (-4.9,-1.7) (w43) {\tiny{$[k+m]$}};
\node[state] at (-4.9,-3) (w44) {\tiny{$[k+3m]$}};

\node[state] at (-6.6,3.2) (w45) {};
\node[state] at (-6.6,2.8) (w46) {};

\node[state] at (-6.6,1.9) (w47) {};
\node[state] at (-6.6,1.5) (w48) {};

\node[state] at (-6.6,1.2) (w49) {};
\node[state] at (-6.6,0.8) (w50) {};

\node[state] at (-6.6,0.6) (w51) {};
\node[state] at (-6.6,0.2) (w52) {};

\node[state] at (-6.6,0.1) (w53) {};
\node[state] at (-6.6,-0.2) (w54) {};

\node[state] at (-6.6,-0.8) (w55) {};
\node[state] at (-6.6,-1.2) (w56) {};

\node[state] at (-6.6,-1.5) (w57) {};
\node[state] at (-6.6,-1.9) (w58) {};

\node[state] at (-6.6,-2.8) (w59) {};
\node[state] at (-6.6,-3.2) (w60) {};


\draw[->, label,red] (w9) -- (w3) ;
\draw[->, label,red] (w10) -- (w3);

\draw[->, label,red] (w11) -- (w4) ;
\draw[->, label,red] (w12) -- (w4);

 \draw[->, label,blue] (w37) -- (w9) ;
  \draw[->, label,blue] (w38) -- (w9);

   \draw[->, label,blue] (w39) -- (w10);
  \draw[->, label,blue] (w40) -- (w10);

  \draw[->, label,blue] (w41) -- (w11);
  \draw[->, label,blue] (w42) -- (w11);

  \draw[->, label,blue] (w43) -- (w12);
  \draw[->, label,blue] (w44) -- (w12);

   \draw[->, label,red,dashed] (w45) -- (w37) ;
  \draw[->, label,red,dashed] (w46) -- (w37);

  \draw[->, label,red,dashed] (w47) -- (w38) ;
  \draw[->, label,red,dashed] (w48) -- (w38);

  \draw[->, label,red,dashed] (w49) -- (w39) ;
  \draw[->, label,red,dashed] (w50) -- (w39);

\draw[->, label,red,dashed] (w51) -- (w40) ;
  \draw[->, label,red,dashed] (w51) -- (w40);

  \draw[->, label,red,dashed] (w53) -- (w41) ;
  \draw[->, label,red,dashed] (w54) -- (w41);

\draw[->, label,red,dashed] (w55) -- (w42) ;
  \draw[->, label,red,dashed] (w56) -- (w42);

  \draw[->, label,red,dashed] (w57) -- (w43) ;
  \draw[->, label,red,dashed] (w58) -- (w43);

   \draw[->, label,red,dashed] (w59) -- (w44) ;
  \draw[->, label,red,dashed] (w60) -- (w44);

  \end{tikzpicture}
    }
    \caption{A $k$-centered butterfly (Right), and its body (Left). The red and blue arrows represent Roman's and Columba's plausibility orderings, respectively.}
    \label{fig:butterfly}
\end{figure}

\noindent \textbf{Step 1: Body of the butterfly.} The \emph{body} $\mathcal{M}_k^0$ of the butterfly, illustrated in Figure~\ref{fig:butterfly}, Left, is defined as:
    \begin{itemize}
        \item $W^0 = \{w_0, w_1, w_2, w_3, w_4\}$.
        \item $w_0 >^0_\rho w_1$, $w_0 >^0_\rho w_2$, $w_0 >^0_\chi w_3$, $w_0 >^0_\chi w_4$ 
        \item $V^0(k) = \{w_0\}, V^0(k-m) = \{w_1, w_3\}, V^0(k+m) = \{w_2, w_4\}$, $V^0(\top) = W^0$, and $V^0([l]) = \emptyset$ for all $l \neq k, k-m, k+m$.
        \end{itemize}
The world $w_0$ in the body of the butterfly centered on $k$ is called its \textit{center}. The body of the butterfly models the shared environment, i.e. the reflexive common indicator $A$ in our philosophical argument, of Roman and Columba. As we shall see below, taking $k=300$ ensures that while both know that the mast is at most $300+m$ and at least $300 -m$ tall, they have an iterated reason to believe that it is 300 cm, that is, $r^n([300])$ holds and therefore also that the mast is larger than 100 cm.  

\noindent \textbf{Step 2: The $k$-centered butterfly.}
Suppose $\mathcal{M}_k^n$ is defined. We call a state $w$ minimal in $\mathcal{M}^n_k$ whenever there is no state $w'$ such that $w >_i w'$ for some agent $i$. Observe that if $w$ is minimal in $\mathcal{M}_k^n$ then there is a state $w'$ such that $w' >_i w$ for $i \in \{\rho, \chi\}$. It also holds that for each such minimal state $w$, it is never the case that there are states $w'$ and $w''$ such that $w' >_\rho w$ and $w'' >_\chi w$. In other words, a minimal state $w$ is connected to other states $w'$ by the relation of a unique agent $i \in \{\rho, \chi\}$. For that reason, we can label minimal states with that unique agent $i \in \{\rho, \chi\}$, and so call them $i$-terminating. $\mathcal{M}_k^{n+1}$ is then constructed as follows. For each $i$-terminating state $w$ in $\mathcal{M}_k^n$, we add two additional states $v$ and $v'$ to $W^n$, and extend the relations $\geq^n_i$ and the valuation $V^n$ as follows. 
    \begin{itemize}
        \item Add all pairs $w >^{n+1}_j v$ and $w >^{n+1}_j v'$, with $j \neq i$, to the relevant $\geq^n_j$.
        \item Extend the valuation $V^n$ with: $V^{n+1}([\bar{V}^{n}(w)+m)]) = \{v\}$,  $V^{n+1}([\bar{V}^{n}(w)-m])) = \{v'\}$, $V^n(\top)= W^{n+1}$.
    \end{itemize}
    In this definition $\bar{V}^{n}(w)$ is the inverse of the valuation function $V^n$, giving us the number $k \in \mathbb{N}$ that is assigned to $w$ in $\mathcal{M}_k^n$.  If $\bar{V}^{n}(w)-m < 0$, we only add $v$. $\mathcal{M}_k$, the butterfly centered on $k$, is then defined as $\bigcup_{n\in \mathbb{N}} \mathcal{M}_k^n$. See Figure~\ref{fig:butterfly}, Right.

The induction steps thus extend the butterfly with two ``wings", whose function is to capture the higher-order uncertainty of Roman and Columba. Indeed, as we shall also see below, moving along each of the paths starting at $k$ generates a chain of ``Roman/Columba considers it possible that Columba/Roman considers it possible... that the mast is $l$ cm", even for $l$ smaller than 100.  

\noindent \textbf{Step 3: The butterfly flutter.} The last step is simply to put together the $k$-centered butterflies, for all $k \in \mathbb{N}$, to build a flutter. These copies will be used to define the conditional beliefs. A \emph{butterfly flutter} $\mathcal{S}$ is the disjoint union $\DUnion_{k \in \mathbb{N}} \mathcal{M}_k$ of the $k$-centered butterflies for all $k \in \mathbb{N}$, together with a selection function $f$. 

We can now define a selection function in detail as follows: For all $w$ and $[k] \in Prop$, $f([k], w)$ is the center of the $k$-centered butterfly in $\mathcal{S}$. For $\top$, we set $f(\top, w)$ to be the center of the $k$-centered butterfly for the unique $k$ such that $w \in V([k])$. Observe that by this construction all conditional and unconditional belief statements are evaluated at the center of a butterfly. Introducing a selection function as such allows us to evaluate within epistemic plausibility models reasons to believe a proposition $A$ that is consistent with considering $\neg A$ possible.

\noindent \textbf{Key facts:} Take the butterfly flutter defined above, and $w$ be the center of the butterfly centered on $300$, and let $m$ be the margin of error. The following facts capture the key points of our philosophical argument. Their proofs are elementary, so for reason of space we only sketch them here. 
\begin{enumerate}
    \item Both agents know that the mast is $300 \pm m$ cm tall: $\mathcal{S}, w \models K_\rho ([300 -m ] \vee [300] \vee [300+m])$, and similarly for $\chi$. Argument: clear by construction. 
    \item Both agents have an unconditional reason to believe that the mast is $300$ cm tall: $\mathcal{S}, w \models R_\rho ([300])$, and similarly for $\chi$. Argument: $[300]$ is true at $w$, by the definition of $f$, we have $f(\top,w)=w$,  and $w$ is maximal in $[w]_\rho$.
    \item Roman and Columba have iterated higher-order reasons to believe that the mast is $300 \pm m$ cm tall: $\mathcal{S}, w \models r^n_{\{\rho,\chi\}} ([300 -m ] \vee [300] \vee [300+m])$. This also means $\mathcal{S}, w \models r^n_{\{\rho,\chi\}} ([> 100])$. Argument: We have already seen that $f(\top,w)=w$ and that $w$ is maximal in $[w]_\rho$ and $[w]_\chi$. So at every iteration of higher-order reasons to believe the only state considered is $w$ itself.
    \item Roman and Columba have the corresponding counterfactual, iterated higher-order reasons to believe that the mast is $k$ cm tall for any $k \in \mathbb{N}$: For $i \in \{\rho, \chi\}$, we have $\mathcal{S}, w \models r^n(R_j([k])|[k])$. Argument: This is ensured by modifying for $k$ the argument above for 300, together with the fact that $f(k, w)$ is the center of the relevant butterfly centered on $k$ in $\mathcal{S}$. 
    \item  $[>100]$ is not common knowledge in the iterated sense, i.e., for all $l$ such that $300-(l.m) > 0$, there is $r$ such that $\mathcal{S}, w \models \langle K_i \rangle \langle K_j^r \rangle [300-(l.m)]$. Argument: recall that $v \approx_i v'$ is defined as $v \geq_i v'$ or $v' \geq_i v$. That the fact holds can then be verified by following the uppermost paths to the left and the right of the center in Figure~\ref{fig:butterfly}, assuming that $k=300$. 
\end{enumerate}
Facts 1 to 5 together flesh out the philosophical argument that we developed in Section~\ref{sec:philarg}. They show that iterative higher-order reasons to believe that $[>100]$ is consistent, now in the precise, logical sense of the term, with, first, a failure of iterative common knowledge of that same proposition and, second, the relevant iterative, higher-order counterfactual reasons to believe that we described in Section~\ref{sec:philarg}. The fact that these counterfactual reasons to believe are consistent with $[>100]$ at $w$ shows that they do not detach.

\section{Conclusion}\label{sec:conclusion}
We have argued both philosophically and formally that understanding publicity in terms of Lewisian common knowledge allows one to give a satisfactory analysis of the Sailboat example. The crux of the philosophical argument is the fact that having a reason to believe in a proposition $A$ is consistent with considering it possible that $A$ is false. This allows Lewisian common knowledge of $[>100]$ to hold even when iterative common knowledge of that proposition fails. The formal construction in Section~\ref{sec:epm} makes this intuition concrete by modeling, respectively, reasons to believe using plausibility orderings and knowledge using their connected components.  

We should emphasize that these arguments should not be read as criticism of, or proposal to ``replace" the iterative or fixed-point definitions of common knowledge with its Lewisian counterpart. These definitions have been extremely helpful in a number of areas, in particular for understanding the type of ``back-and-forth" reasoning underlying game-theoretical solution concepts~\cite{perea2012epistemic,sep-epistemic-game}. What the argument suggests is that Lewisian common knowledge is perhaps better suited than iterative common knowledge to capture precisely the pre-theoretical notion of publicity, and that this should be taken into account when analyzing concrete cases that involve the latter. Of course, there remain numerous other cases where this view should be put to test: e.g. the classical Byzantine generals example, the Halpern-Moses problem, the consecutive numbers example, or philosophical accounts of collective agency that rely on iterative common knowledge. Other logical framework could be used to represent those cases, for instance the ones developed in~\cite{baltag2018some} or in~\cite{van2021wanted,bilkova2024bisimulation}  We leave those for future work, but conjecture that Lewisian common knowledge would prove illuminating in those cases too. 

\section*{Acknowledgements} The authors would like to thank the anonymous reviewers of TARK for helpful comments and suggestions, as well as the participants of the CELIA project meeting 2025. Financial support of the DFG-GACR project CELIA (RO 4548/13-1) is gratefully acknowledged. 

\bibliographystyle{eptcs}
\bibliography{bib}

\end{document}